\documentclass[traditabstract]{aa}

\usepackage{graphicx,natbib,txfonts}

\newcommand{\ii}{\mathrm{i}}
\renewcommand{\d}{\mathrm{d}}

\begin{document}

\title{Calibration biases in measurements of weak lensing}
\author
 {Matthias Bartelmann\inst{1} \and
  Massimo Viola\inst{1} \and
  Peter Melchior\inst{1, 3, 4} \and
  Bj\"orn M.~Sch\"afer\inst{2}}
\institute
 {Zentrum f\"ur Astronomie der Universit\"at Heidelberg, ITA, Albert-Ueberle-Str.~2, 69120 Heidelberg, Germany, \email{mbartelmann@ita.uni-heidelberg.de} \and
 Zentrum f\"ur Astronomie der Universit\"at Heidelberg, ARI, M\"onchhofstr.~12--14, 69120 Heidelberg, Germany \and
 Center for Cosmology and Astro-Particle Physics, The Ohio State University, 191 W. Woodruff Ave., Columbus, Ohio 43210, USA \and
 Department of Physics, The Ohio State University, 191 W. Woodruff Ave., Columbus, Ohio 43210, USA}

\date{\today}

\abstract{As recently shown by \citeauthor{2010MNRAS.tmp.1576V}, the common (KSB) method for measuring weak gravitational shear creates a non-linear relation between the measured and the true shear of objects. We investigate here what effect such a non-linear calibration relation may have on cosmological parameter estimates from weak lensing if a simpler, linear calibration relation is assumed. We show that the non-linear relation introduces a bias in the shear-correlation amplitude and thus a bias in the cosmological parameters $\Omega_\mathrm{m0}$ and $\sigma_8$. Its direction and magnitude depends on whether the point-spread function is narrow or wide compared to the galaxy images from which the shear is measured. Substantial over- or underestimates of the cosmological parameters are equally possible, depending also on the variant of the KSB method. Our results show that for trustable cosmological-parameter estimates from measurements of weak lensing, one must verify that the method employed is free from ellipticity-dependent biases or monitor that the calibration relation inferred from simulations is applicable to the survey at hand.}


\maketitle

\section{Introduction}

Measurements of weak gravitational lensing have developed into one of the main diagnostic tools for the dark-matter distribution on large scales. In principle, the ellipticity of the surface-brightness distribution in the images of distant galaxies is quantified in some way, averaged over sufficiently many images, and related to the mean ellipticity expected in presence of gravitational shear to obtain a local estimate of the shear. Even though typical lensing effects are very weak and superposed on a substantial shape-noise contribution from the galaxies, many measurements have succeeded and routinely produce cosmological parameter estimates which are typically well in agreement with alternative determinations, or supplementing them in a highly plausible way. See \cite{2010CQGra..27w3001B, 2001PhR...340..291B} for recent reviews.

There are two ways of quantifying the light distribution of faint galaxies. One of them, a model-free approach, measures sufficiently high-order moments of the surface-brightness distribution and combines them into ellipticity estimates that can then be compared to the shear \citep{1995ApJ...449..460K}. The other compares model images to real data and varies the applied shear until both match optimally \citep{2007MNRAS.382..315M, 2002AJ....123..583B, 1999A&A...352..355K}. We are here concerned with the first approach, which has the advantage of not assuming an intrinsic image shape. \cite{2010MNRAS.404..458V} have shown how too simplistic or rigid model shapes can fundamentally limit the accuracy of shear measurements.

Let $I(\vec\theta)$ be the surface-brightness distribution of a galaxy image, then the moments are defined as
\begin{equation}
 Q_{ij\ldots k} = \int\d^2\theta\,I(\vec\theta)\,W(\theta)\,\theta_i\theta_j\ldots\theta_k\;,
\label{eq:01}
\end{equation} 
with the integral is carried out over the image and the weight function is introduced to cut the integration off in order to limit the inclusion of noise. The three independent second moments $Q_{ij}$ are combined to form the complex ellipticity
\begin{equation}
  \chi = \frac{Q_{11}-Q_{22}+2\ii Q_{12}}{Q_{11}+Q_{22}}\;.
\label{eq:02}
\end{equation}
The intrinsic ellipticity of a source, $\chi^\mathrm{s}$, is related to the ellipticity $\chi$ of a sheared image by
\begin{equation}
  \chi^\mathrm{s} = \frac{\chi-2g+g^2\chi^*}{1+|g|^2-2\Re(g\chi^*)}\;.
\label{eq:03}
\end{equation}
The reduced shear $g=\gamma(1-\kappa)^{-1}$ appears because the shear $\gamma$ itself is measurable only in combination with the convergence $\kappa$. Since $\gamma\in\mathbb{C}$, so is $g$.

Two essential problems arise in any application of Eq.~(\ref{eq:02}) to shear measurements. First, the equation is strictly non-linear. The lowest-order linear approximation,
\begin{equation}
  \chi^\mathrm{s} = \chi-2g\;,
\label{eq:04}
\end{equation}
is typically not sufficiently precise, thus higher-order corrections need to be applied. Second, the observed image is further distorted after being gravitationally sheared. It is convolved with the point-spread function of the optical system and of the atmospheric seeing and it is pixellised by the detector. Moreover, intrinsically highly elliptical sources are less susceptible to the gravitational shear applied. These effects need to be corrected.

The standard procedure for these corrections has been defined by \citeauthor{1995ApJ...449..460K} (\citeyear{1995ApJ...449..460K}, hereafter KSB). Although we adopt KSB here as a specific example against which we can quantify our statements, we emphasise that the central issue of this paper is by no means a criticism of KSB, but the bias caused by any non-linear calibration relation that is approximated by a linear one. The KSB method is particularly relevant in this context because it is very fast and has been shown to perform well for sources with high signal-to-noise \citep{2010MNRAS.405.2044B}. It is routinely being used also in shear analyses of galaxy clusters \cite[see][for a recent overview]{2007MNRAS.379..317H}, where calibration biases are more severe than for cosmic shear.

In a recent analysis of KSB \citep{2010MNRAS.tmp.1576V}, we have identified three potentially problematic steps or assumptions, which are: (1) KSB averages shear measurements from galaxy images, rather than measuring the shear from averaged galaxy images, but the shear measurement and the average do not generally commute. (2) KSB implicitly assumes that galaxy ellipticities are small, while weak gravitational lensing only assures that the change in ellipticity due to the shear is small. (3) KSB approximately corrects for the convolution with the point-spread function (PSF), but does not deconvolve it. Step (1) leads to biased results, while assumptions (2) and (3) partially counteract in a way dependent on the width of the weight function and of the PSF. These effects were analysed in detail in \cite{2010MNRAS.tmp.1576V}. Neither of these assumptions is inherent to the moment-based approach to weak lensing. In fact, we recently proposed a novel method, dubbed DEIMOS \citep{2010arXiv1008.1076M}, which avoids these assumptions and whose performance is therefore much more stable against variations of the PSF shape and the source ellipticity.

Here, we study the implications of ellipticity-dependent biases, by the specific example of KSB, for the constraints on cosmological parameters. Section~2 summarises the shear calibration relations for different variants of KSB. We describe simulations of the resulting biases in Sect.~3 and present our results in Sect.~4.

\section{Shear calibration relations}

The KSB method describes the relation between the ellipticities of source and image by a tensor $P^\mathrm{sh}$,
\begin{equation}
  (\chi-\chi^\mathrm{s})_\alpha = P^\mathrm{sh}_{\alpha\beta}\,g^\beta\;.
\label{eq:05}
\end{equation} 
Ideally, to lowest order, $P^\mathrm{sh}$ is twice the identity, $P^\mathrm{sh}_{\alpha\beta} = 2\delta_{\alpha\beta}$, as Eq.~(\ref{eq:04}) shows. The non-linearity of Eq.~(\ref{eq:03}), the presence of a weight function $W$ in Eq.~(\ref{eq:01}) and the convolution by the point-spread function complicate matters considerably. Actual implementations of the KSB method differ in the approximations and assumptions made in the concrete representation of the tensor $P^\mathrm{sh}$. As in \cite{2010MNRAS.tmp.1576V}, we investigate four of them here:
\begin{itemize}
\item The original KSB method, labelled \textit{KSB}, which treats the order of the reduced shear and the image ellipticity inconsistently;
\item A simplification of KSB, labelled \textit{KSB-1}, which stays consistently at the linear order;
\item A common modification of KSB, labelled \textit{KSB-tr}, which approximates the tensor $P^\mathrm{sh}$ by
\begin{equation}
  P^\mathrm{sh}_{\alpha\beta} =
  \frac{1}{2}\mathrm{tr}\left(P^\mathrm{sh}\right)\delta_{\alpha\beta}\;,
\label{eq:07}
\end{equation}
which lacks mathematical justification but performs well in practice \cite[see][for an application to the COSMOS field with its narrow PSF]{2010A&A...516A..63S};
\item A consistent extension of KSB to third order in shear and ellipticity, labelled \textit{KSB-3}, which we developed in \cite{2010MNRAS.tmp.1576V}.
\end{itemize}

\begin{figure}[ht]
  \includegraphics[width=\hsize]{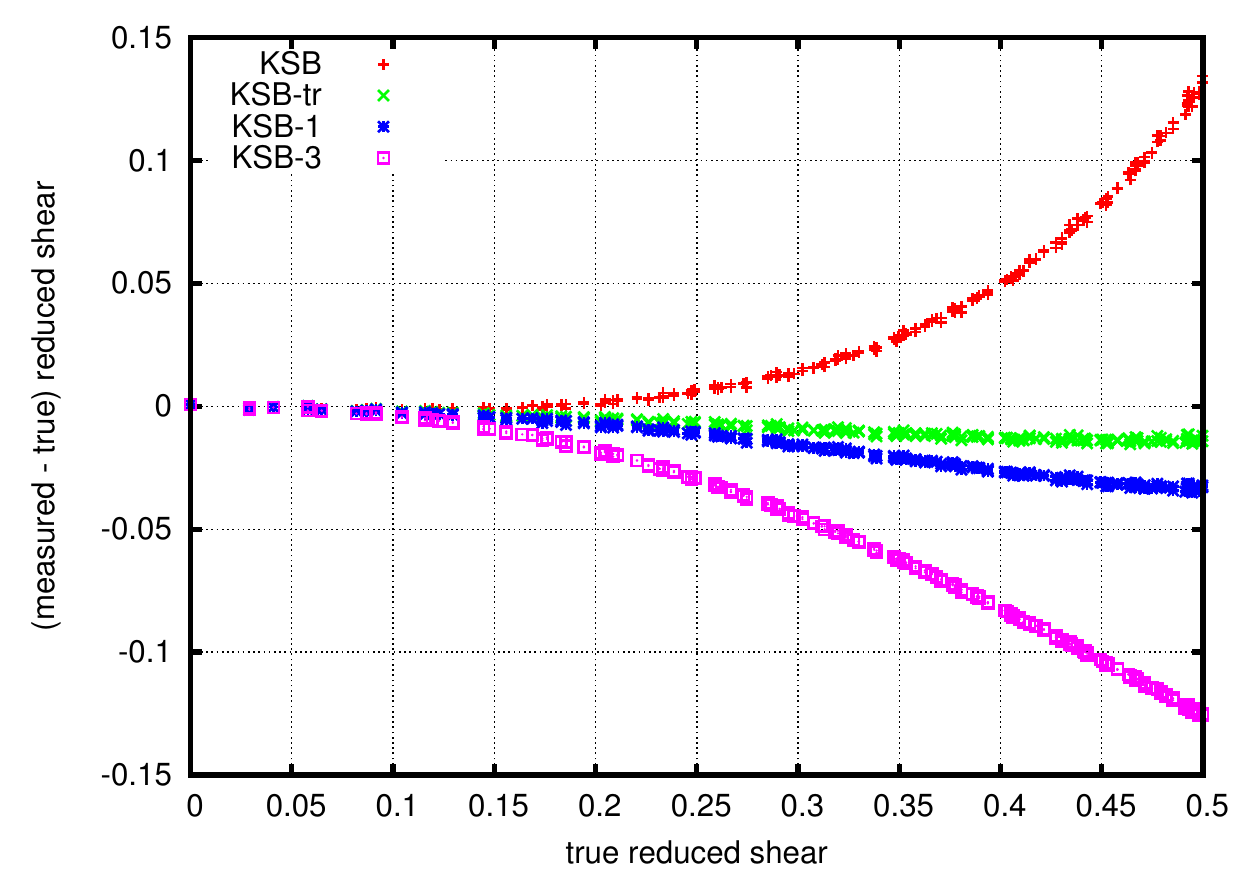}\\
  \includegraphics[width=\hsize]{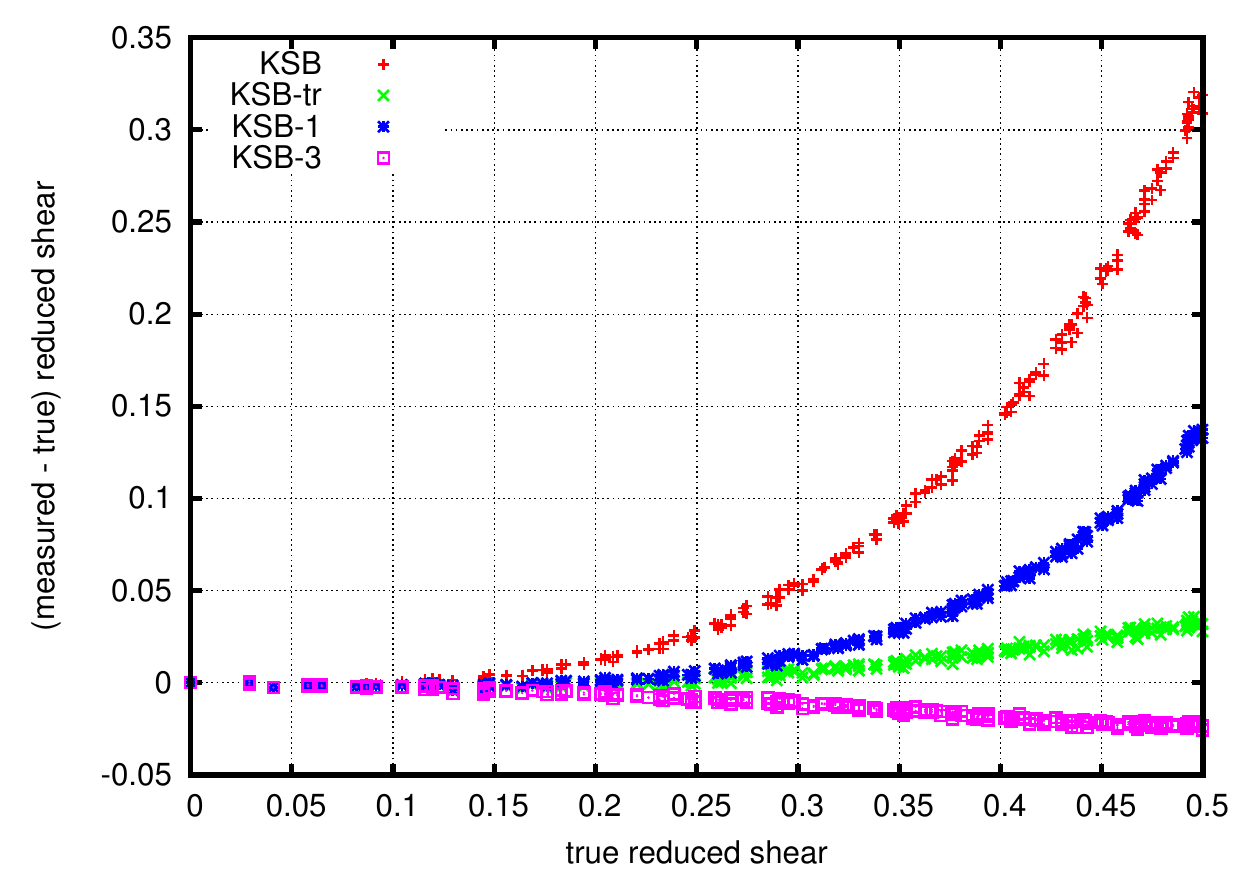}
\caption{The difference between the measured reduced-shear estimate and the true reduced shear is shown as a function of the true reduced shear, for different measurement methods as indicated by the symbols. The ideal case of a measurement obtaining the true value exactly corresponds to the horizontal zero line. \textit{Top panel}: ground-based observations with a wide point-spread function. \textit{Bottom panel}: space-borne observations with a point-spread function of negligible width.}
\label{fig:1}
\end{figure}

We apply these methods of shear measurement to simulated noise-free galaxy images. This proceeds along the following steps:
\begin{itemize}
\item A value for the reduced shear $g=g_1+\ii g_2$ is randomly drawn from a flat distribution for the $g_i$ between $[0, 0.1]$.
\item A value for the intrinsic ellipticity $\chi^\mathrm{s}$ is drawn such that its modulus $|\chi^\mathrm{s}|=0.3$ and its orientation is random within $[0, 2\pi)$. For simplicity, we ignore an intrinsic ellipticity dispersion.
\item Each galaxy is sheared with the reduced shear $g$. This is repeated with $N\approx100$ galaxies to suppress the uncertainty due to shape noise in the shear estimate.
\item All galaxies are convolved with a given PSF with a Moffat profile. The size of the weight function is set equal to the size of the convolved galaxy image.
\item The ellipticity of each image $\chi^\mathrm{obs}$ is measured. From each measurement, a shear estimate $\tilde g$ is computed with each of the four variants of the KSB method.
\end{itemize}
The results are illustrated in Fig.~\ref{fig:1} for two different choices of the PSF, where the measured reduced-shear estimate $\tilde g$ is plotted against the true reduced shear $g$.

Finally, the shear estimates are averaged, accounting for the shear responsivity, to arrive at the average shear estimate
\begin{equation}
  g^\mathrm{(variant)}(g) = \frac{1}{1-\chi^{\mathrm{s}2}/2}\frac{1}{N}\sum_{i=1}^N\tilde g_i(g)\;.
\label{eq:08}
\end{equation} 

The results shown in the two panels of Fig.\ref{fig:1} are based on a narrow PSF (bottom) mimicking a space-borne observation, and a wide PSF (top) that could be typical for a ground-based observation. Both PSFs are fully characterised by their Moffat exponent $\beta$ and their FWHM. For the narrow PSF, $(\beta, \mathrm{FWHM})=(2, 0.5\,R_\mathrm{e})$; for the wide PSF, $(\beta, \mathrm{FWHM})=(5, 5\,R_\mathrm{e})$, where $R_\mathrm{e}$ is the scale radius of the S{\'e}rsic profile assumed for the galaxy images. The ideal relation (following the diagonal) is marked by a dash-dotted line. These figures illustrate three main points. First, the measured shear estimate falls below or above the true shear, depending on the method used for the conversion of observed ellipticities to shear. Second, the relations between measured shear estimate and the true shear are non-linear, which will turn into our main issue for this paper. Third, the deviation of the shear estimate from the true shear, even its sign, depend on the width of the PSF. While KSB-3, for example, performs almost perfectly for a narrow PSF, it underestimates the true shear if the PSF is wide. The origin of these trends has been identified in \cite{2010MNRAS.tmp.1576V}. Here, we work out the consequences for the cosmological-parameter determinations from such measurements.

\section{Biases on cosmological parameters}

Given the non-linear relations between the shear estimate and the true shear, we now consider the following situation. Suppose a cosmic-shear observation is being carried out, ellipticities are measured and converted to a shear estimate following one of the KSB variants described above, the shear correlation function is measured and cosmological parameters (essentially the matter density $\Omega_\mathrm{m0}$ and the normalisation parameter $\sigma_8$). Suppose further that the shear estimate is in fact calibrated in the analysis process, e.g.~by means of synthetic data, however assuming a \textit{linear} rather than the underlying, non-linear relationship between the shear estimate and the true shear. In other words, the calibration is supposed to be carried out assuming that the response of the measurement to the shear is independent of the shear itself, rather than depending on it. Given such a procedure, which can perhaps be considered close to the common practice of identifying additive and multiplicative shear biases \citep{2006MNRAS.368.1323H}, how biased would the cosmological parameters be, if at all?

In order to answer this question, the following steps need to be carried out:

First, we fix a reference cosmological model and compute the shear correlation function for it. To be specific, we choose the correlation function $\xi_+(\theta)$, which is given by
\begin{equation}
  \xi_+(\theta) = 2\pi\int_0^\infty l\d l\,P_\kappa(l)\,F(l\theta)\;,\quad
  F(x)=\frac{\mathrm{J}_0(x)}{4\pi^2}\;.
\label{eq:09}
\end{equation}
The power spectrum $P_\kappa(l)$ of the convergence is the usual weighted projection of the dark-matter fluctuation power spectrum $P_\delta(k)$ in Limber's approximation. In a spatially-flat universe for sources at a fixed redshift,
\begin{equation}
  P_\kappa(l) = \frac{9}{4}\left(\frac{H_0}{c}\right)^4\Omega_\mathrm{m0}^2
  \int_0^{w_\mathrm{s}}\d w\left(\frac{w_\mathrm{s}-w}{w_\mathrm{s}a(w)}\right)^2
  P_\delta\left(\frac{l}{w}\right)\;,
\label{eq:10}
\end{equation}
where $w$ and $w_\mathrm{s}$ are the comoving angular-diameter distances to the lensing matter fluctuation and to the sources, respectively (see, e.g., \citealt{2010CQGra..27w3001B, 2001PhR...340..291B} for a derivation).

Second, we choose hypothetical survey parameters, define bins $\{\theta_i\}$ for the measurement of the correlation function and compute the covariance matrix $C_{ij}$ for the correlation function $\xi_+$ between these bins. Doing so, we follow the procedure developed by \cite{2008A&A...477...43J}. This allows the computation of two important ingredients. First, we can choose the bins such that the signal-to-noise ratio is approximately the same in each bin. Second, we can calculate the expected uncertainties of the measured correlation function including their mutual correlation between bins. The latter step is conveniently achieved by rotating the bin vector into the principal-axis frame of the covariance matrix, drawing independent Gaussian random numbers with the appropriate variance, and inverting the rotation.

Next, we fit the calibration relations shown in Fig.~\ref{fig:1} with a realistic quadratic dependence
\begin{equation}
  \tilde g = a + bg+cg^2
\label{eq:11}
\end{equation}
and an assumed linear dependence
\begin{equation}
  \tilde g' = a' + b'g\;.
\label{eq:12}
\end{equation}
These correlation functions of the shear estimate and the true shear are then related by
\begin{equation}
  \tilde\xi_+ = a^2+2ac\sigma_g^2+b^2\xi_+\;,\quad
  \tilde\xi_+' = a'^2 + b'^2\xi_+
\label{eq:13}
\end{equation}
and must be corrected accordingly by an offset and an amplitude to arrive at $\xi_+$. The offsets are
\begin{equation}
  \delta\xi_+ = a^2+2ac\sigma_g^2\quad\hbox{and}\quad\delta\xi' = a'^2
\label{eq:14}
\end{equation}
for the quadratic and linear relations, while the amplitudes are $b^2$ and $b'^2$, respectively. We shall assume in the following that the offset can be faithfully corrected in any survey by demanding that the correlation function approach zero at large angles. However, the correction of the amplitude remains. Our main issue for the following discussion is that the amplitude $b'^2$ derived from the assumed linear calibration relation differs from the amplitude $b^2$ expected from the realistic, quadratic calibration relation. We thus assume that the measured shear correlation is corrected by $b'^{-2}$ while it should be corrected by $b^{-2}$. That is, we erroneously infer the correlation function
\begin{equation}
  \xi_+' = \frac{b^2}{b'^2}\,\xi_+
\label{eq:15}
\end{equation}
instead of $\xi_+$, which is then compared to theory and gives rise to a bias in the cosmological parameters.

Our third and final step is thus quite simple: We take the simulated correlation function including its modelled uncertainty and error bars, multiply it by the factor $b^2/b'^2$, and then fit theoretical correlation functions to them minimising the $\chi^2$ function
\begin{equation}
  \chi^2 = \left[
    \xi_{+, i}'-\xi_+(\theta_i, p)\right]C_{ij}^{-1}\left[\xi_{+, j}'-\xi_+(\theta_j, p)
  \right]\;,
\label{eq:16}
\end{equation} 
where $p$ symbolically abbreviates all parameters entering the calculation of the theoretically expected correlation function $\xi_+$.

\begin{figure*}[ht]
  \includegraphics[width=\hsize]{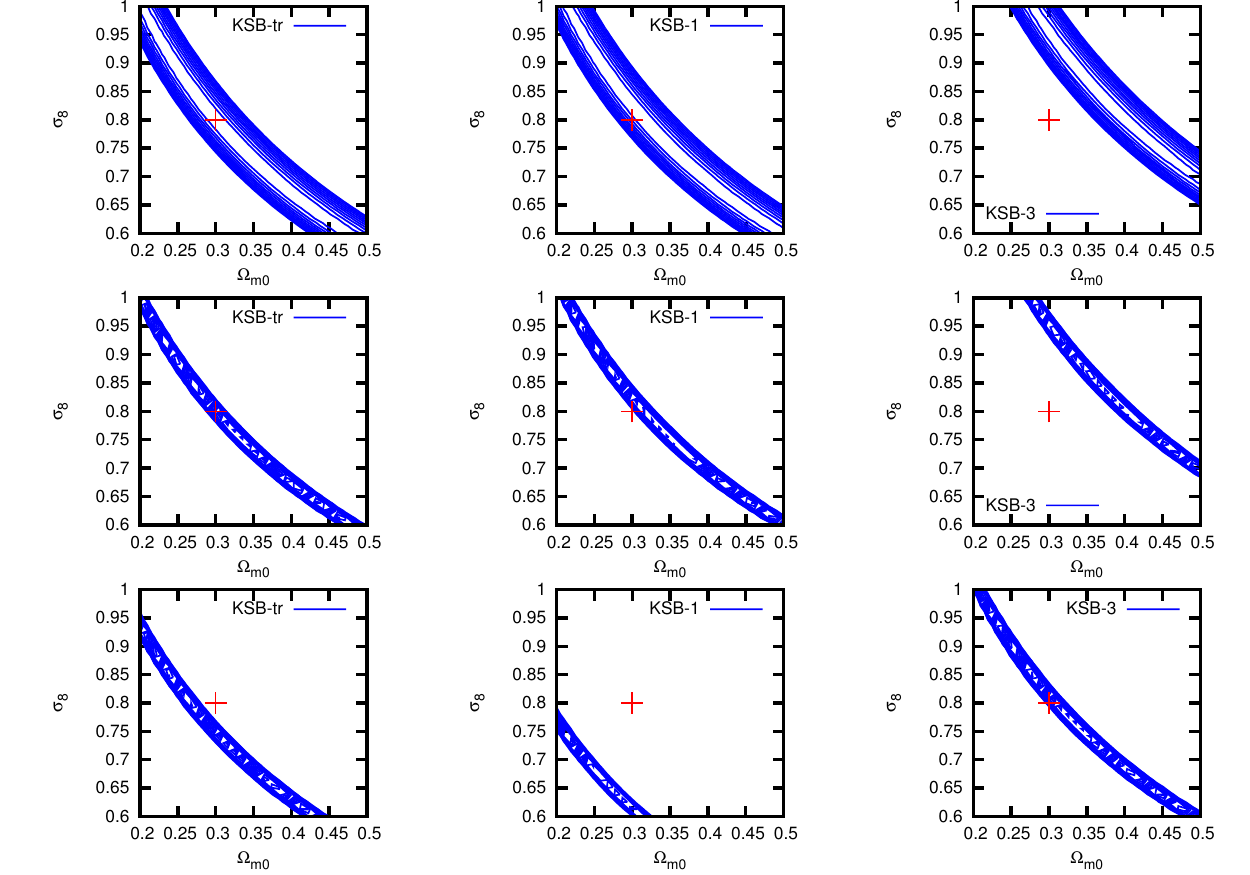}
\caption{Likelihood contours are shown in the $\Omega_\mathrm{m0}$-$\sigma_8$ plane for: A CFHT-like survey in the top row and a wide survey in the middle and bottom rows; three variants of the KSB method, i.e.~KSB-tr in the left, KSB-1 in the middle and KSB-3 in the right columns; a wide PSF in the top and middle rows and a narrow PSF in the bottom row. The Figure shows that cosmological parameters tend to be biased high for wide and low for narrow PSFs, and that the magnitude of the bias depends on the variant of KSB chosen.}
\label{fig:2}
\end{figure*}

The $\chi^2$ contours following from this procedure are shown for three variants of KSB in Fig.~\ref{fig:2}.

\section{Results and conclusion}

The $\chi^2$ contours shown in these Figures were calculated assuming two different types of survey, whose parameters control the covariance matrices $C_{ij}$. The assumed survey parameters are listed in Tab.~\ref{tab:1}. The cosmological parameters are set to $\Omega_\mathrm{m0}=0.3$, $\Omega_{\Lambda0}=1-\Omega_\mathrm{m0}$, $h=0.7$ and $\sigma_8=0.8$ in both cases.

\begin{table}[ht]
\begin{center}
\begin{tabular}{lrr}
Name & CFHT & wide \\
\hline
$A$ & 34.2 & 100.0 \\
$n_\mathrm{gal}$ & 13.3 & 30.0 \\
$\sigma_\epsilon$ & 0.42 & 0.30 \\
$\bar z_\mathrm{s}$ & 0.9 & 1.0 \\
\end{tabular}
\end{center}
\caption{Assumed parameters for a CFHT-like and a wider weak-lensing survey. $A$ is the area in square degrees, $n_\mathrm{gal}$ the number density of background galaxies per square arc minute, $\sigma_\epsilon$ is the variance of the intrinsic ellipticity, and $\bar z_\mathrm{s}$ is the mean source redshift.}
\label{tab:1}
\end{table}

We leave out the original version of KSB because most of its likelihood contours fall off the parameter region shown. This is because the calibration relation of the original KSB method has the largest non-linear contribution and thus produces the strongest bias. The main pieces of information evident from Fig.~\ref{fig:2} are:
\begin{itemize}
\item The wide survey (results shown in the lower two rows) narrows the contours compared to the CFHT-like survey, but does otherwise not affect the results.
\item For a wide PSF (results shown in the upper two rows), cosmological parameters inferred from the likelihood contours are biased high for the first- and third-order KSB variants (second and third row), while KSB-tr is almost perfectly unbiased even with the narrow contours of the wide survey. The bias is mild for KSB-1 and substantial for KSB-3. This reflects the shape of the calibration curves for the wide PSF as shown in the lower panel of Fig.~\ref{fig:1}. Since they are convex from above, a linear fit to them overestimates the shear and thus leads to an overestimate of the shear correlation amplitude.
\item For a narrow PSF (results shown in the lower row), the inferred cosmological parameters are biased low for KSB-tr and KSB-1, while they are almost unbiased for KSB-3. The explanation is similar as above, since now the calibration relations are convex from below, causing linear fits to underestimate the shear.
\end{itemize}
Of course, the amount and the direction of the bias depend on the exact parameters chosen, in particular for the size of the PSF relative to that of the window function, which is in turn adapted to the object size. The main conclusions we draw from the results in Fig.~\ref{fig:2} are instead:
\begin{itemize}
\item The fact that the response of the KSB method to the shear depends on the shear itself, leading to a non-linear calibration relation between the shear estimate and the true shear, gives rise to possibly substantial biases in cosmological-parameter estimates if a linear calibration relation is assumed.
\item The success of different variants of KSB depends on the width of the PSF, and thus on the circumstances of the survey. Narrow PSFs tend to cause overestimates which wide PSFs tend to cause underestimates.
\item Good results of one variant of KSB, such as KSB-tr for a wide PSF, do not allow to conclude that this variant performs well under all circumstances.
\end{itemize}

The non-linear calibration relation discussed here was not detected by the Shear Testing Programme \cite[STEP,][]{2006MNRAS.368.1323H} because of its restriction to weak shear.

\begin{figure}
  \includegraphics[width=\hsize]{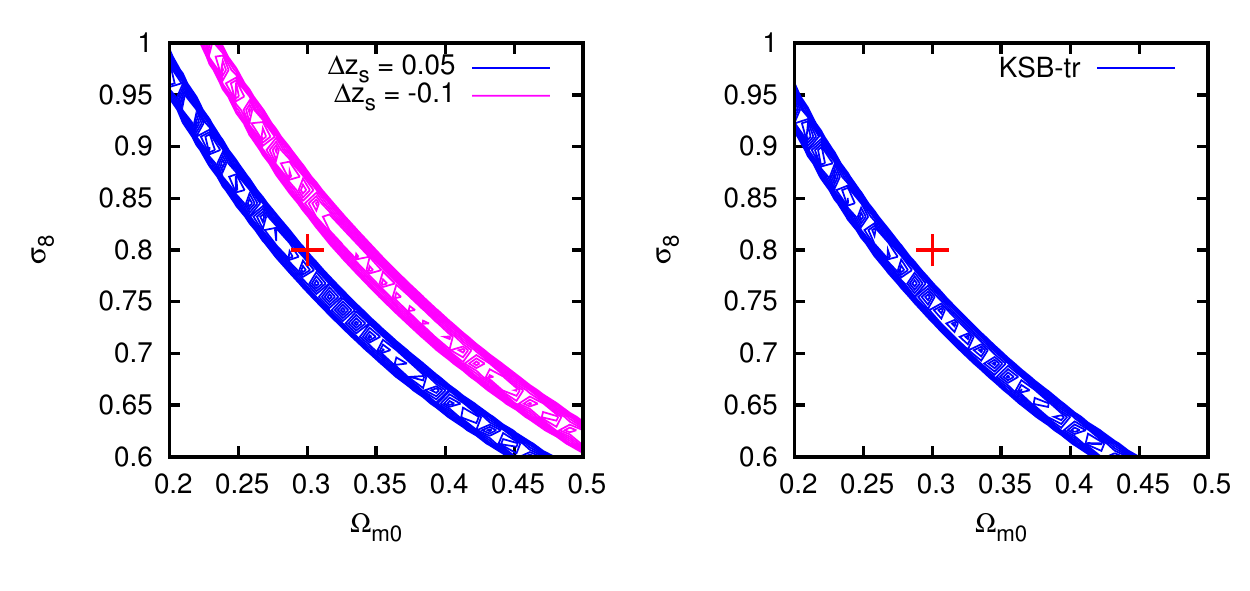}
\caption{Comparison between the likelihood contours in the $\Omega_{m0}$-$\sigma_8$ plane for two different types of bias. \textit{Left panel}: No bias due to shear calibration was assumed, but the source-redshift estimate was assumed to be off by $\Delta z_\mathrm{s}=0.05$ and $\Delta z_\mathrm{s}=-0.1$, as indicated by the line type. The \textit{right panel} is the same as the bottom-left panel in Fig.~\ref{fig:2}. A wide survey was assumed for both panels. The comparison shows that, for a wide survey with narrow PSF, the calibration bias can be comparable or larger than a realistic bias due to an erroneous source redshift.}
\label{fig:3}
\end{figure}

Compared to other potential biases in cosmological constraints from weak lensing, the calibration bias can be substantial. Figure~\ref{fig:3} opposes likelihood contours in the $\Omega_{m0}$-$\sigma_8$ plane for a wide survey, in the left panel without calibration bias in the shear measurement but a bias in the source redshift, and in the right panel without redshift bias but with the calibration bias of the KSB-tr method. The comparison shows that the calibration bias can be as large as a bias to a source redshift estimate off by $\Delta z\approx-0.1$.

Our main concern is not to criticise KSB or its variants, but to emphasise that KSB, in all of its variants, causes a non-linear relation between the shear estimate and the reduced shear. For some variants, the non-linearity is stronger or weaker, depending on circumstances, and some variants are mathematically more consistent than others. Our main conclusion, instead, fits into one statement: Without taking account of its non-linear calibration relation, KSB tends to give biased results. A new method avoiding these problems, called DEIMOS, was proposed by \cite{2010arXiv1008.1076M}.

Other methods to estimate the shear may show similar behaviour. Users concerned with highly accurate shear estimation should therefore investigate whether the method employed exhibits an ellipticity-dependent bias, and if so, whether the shape and parameters of the inferred calibration relation are applicable to the survey at hand. This requires that the simulation, from which the calibration is to be inferred, mimics the actual survey data sufficiently closely, in particular with respect to PSF width and the distributions of galaxy size and ellipticity as well as their variations with redshift and signal-to-noise ratio.

\begin{acknowledgements}
We gratefully acknowledge helpul and improving comments by Catherine Heymans and Thomas Erben. This work was supported in part by the Schwerpunktprogramm 1177 and the Transregio-SFB 33 of the Deutsche Forschungsgemeinschaft, the European RTN ``DUEL'' and the Heidelberg Graduate School of Fundamental Physics. PM was supported by the U.S.\ Department of Energy under Contract No.~DE-FG02-91ER40690.
\end{acknowledgements}


\end{document}